\documentclass[aps,prl,preprint,groupedaddress]{revtex4}
\usepackage{epsfig}
\begin{document}


\title{On the necessity to include event-by-event
fluctuations in experimental evaluation  
of elliptical flow}


\author{R.Andrade, F.Grassi and Y.Hama}
\affiliation{Instituto de F\'{\i}sica-Universidade de S\~ao Paulo}
\author{T.Kodama}
\affiliation{Instituto de F\'{\i}sica-Universidade Federal do Rio de Janeiro}
\author{O.Socolowski Jr.}
\affiliation{CTA/ITA}

\date{\today}

\begin{abstract}
Elliptic flow at RHIC is computed event-by-event with NeXSPheRIO. We 
show that when symmetry of the particle distribution in relation to 
the reaction plane is assumed, as usually done in the experimental 
extraction of elliptic flow, there is  a disagreement between the 
true  and  reconstructed elliptic flows (15-30\% for $\eta$=0, 30\% 
for $p_\perp$=0.5 GeV). 
We suggest a possible
way to take into account the asymmetry
 and get good agreement between
these elliptic flows.
\end{abstract}


\maketitle


Hydrodynamics is one of the main tools to study the collective flow in 
high-energy nuclear collisions. 
Here we discuss results on elliptic flow obtained with the hydrodynamical code NeXSPheRIO. It is a junction of two codes: NeXus and SPheRIO.
The SPheRIO code   is used to compute the hydrodynamical evolution. It is based on
Smoothed Particle Hydrodynamics, a method
originally developed in astrophysics
and adapted to relativistic heavy ion
collisions \cite{spherio}.
Its main advantage is that
 any geometry in the initial conditions can be incorporated.
The  NeXus code  is used to compute the initial conditions
$T_{\mu \nu}$, $j^{\mu}$ and $u^{\mu}$ on a proper time hypersurface \cite{IC}.
NeXSPheRIO is run many times, corresponding to many different events or initial conditions. At the end,
 an average over  final results is performed.
This mimicks experimental conditions.
This is different from the canonical approach in hydrodynamics where
initial conditions are adjusted to reproduce some selected data and are very 
smooth.
This code has been used to study a range of problems concerning relativistic nuclear collisions: effect of fluctuating initial conditions on particle 
distributions \cite{FIC}, energy dependence of the kaon effective temperature 
\cite{kaon}, interferometry at RHIC \cite{HBT}, transverse mass distributions 
at SPS for strange and non-strange particles \cite{strange}, 
effect of the different theoretical and experimental binnings \cite{BJP},
effect of the nature of the quark-hadron transition and of the particle emission mechanism \cite{QM05}.
Here we study the evaluation of elliptic flow using the so-called 
standard method.
The version of NeXSPheRIO used here
 has a first order  quark-hadron transition,
sudden freeze out and no strangeness conservation. 
The only parameter, the freeze out temperature, was assumed to be 150 MeV, since this gives good agreement for the charged particle 
pseudo-rapidity and transverse momentum distributions for all PHOBOS
centrality windows.

Theoretically, the impact parameter $\vec{b}$ is known
and varies in the range of the centrality window chosen. The theoretical,
or true,  elliptic flow parameter
at a given pseudo-rapidity $\eta$
is defined as 
\begin{equation}
<v_2^b(\eta)>=<\frac{\int d^2N/d\phi d\eta \cos[2(\phi-\phi_b)]\, d\phi}
{\int d^2N/d\phi d\eta \, d\phi}>
\end{equation}
$\phi_b$ is the angle between $\vec{b}$ and some fixed reference axis.
The average is performed over all events in the centrality bin.

Experimentally, the impact parameter angle $\phi_b$ is not known. An 
approximation,
$\psi_2$, is estimated. Elliptic flow parameter with respect to this angle, $v_2^{obs}(\eta)$,
 is calculated. Then
a correction is applied to $v_2^{obs}(\eta)$
to account for the reaction plane resolution, leading to 
the experimentally reconstructed elliptic flow parameter $v_2^{rec}(\eta)$.
For example in a Phobos-like way \cite{phobos}
\begin{equation}
<v_2^{rec}(\eta)>=<\frac{v_2^{obs}(\eta)}
         {\sqrt{<\cos[2(\psi_2^{<0}-\psi_2^{>0})]>}}>
\end{equation}
where
\begin{equation}
v_2^{obs}(\eta)=\frac{\sum_i d^2N/d\phi_i d\eta \cos[2(\phi_i-\psi_2)]}
          {\sum_i d^2N/d\phi_i d\eta}
\end{equation}
and
\begin{equation}
\psi_2=\frac{1}{2} \tan^{-1} \frac{\sum_i \sin 2 \phi_i}{\sum_i \cos 2 \phi_i}
\end{equation}

In the hit-based method, 
$\psi_2^{<0}$ and $\psi_2^{>0}$ are determined for subevents 
$\eta < 0$ and $>0$ respectively and
if $v_2$ is computed for a positive (negative) $\eta$,
the sums in $\psi_2$, eq. (4),  are over particles with $\eta < 0$ ($\eta > 0$).
In the track-based method, 
$\psi_2^{<0}$ and $\psi_2^{>0}$ are determined for subevents 
$2.05<\mid \eta \mid < 3.2$,  the sums in $\psi_2$, eq. (4),  
are over particles in both sub-events, 
$v_2$ is obtained for particles around $0<\eta < 1.8$ and reflected
(to account for the different multiplicities between a subevent
and  the sums in eq. (4),
there is also an additional $\sqrt{2\alpha}$ with $\alpha\sim 1$,
in the reaction plane correction
in eq. (2)).
Since both methods are in agreement but only the hit-based method covers a 
large pseudo-rapidity interval, we use this latter method.

We want to check whether the theoretical and experimental estimates are in agreement, i.e.,
$<v_2^b(\eta)>=<v_2^{rec}(\eta)>$. A necessary condition for this, from eq. (2), is,
$<v_2^b(\eta)>\geq <v_2^{obs}(\eta)>$.
In figure 1, we show  the results for $<v_2^b(\eta)>$ (solid line)
and $<v_2^{obs}(\eta)>$ (dashed line).
We see that $<v_2^b(\eta)>\le <v_2^{obs}(\eta)>$ for most $\eta$'s.
So, as shown also in the figure,  dividing  by a cosine to get $<v_2^{rec}(\eta)>$
(dotted curve)
makes the disagreement worse:  $<v_2^{b}(\eta)>$ and 
$<v_2^{rec}(\eta)>$ {\em are} different.
This is true for all three Phobos centrality windows and more pronounced in the most central window.
\begin{center}
\begin{figure}
\epsfig{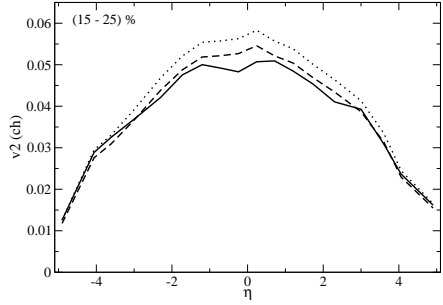}
\caption[]{Comparison between various ways of computing $v_2$
using NeXSPheRIO
for  Phobos 15-25\% centrality window\cite{phobos}:  solid line is 
$v_2^b$,
obtained using 
the known impact parameter angle $\phi_b$, dashed (dotted) line is $v_2^{obs}$
($v_2^{rec}$),
obtained
using the 
reconstructed impact parameter angle $\psi_2$ without (with) reaction plane correction.}  
\end{figure}
\end{center}

Since the standard way to 
include the correction for the reaction plane resolution (eq. (2))
seems inapplicable,
 we need to understand why.
When we look at the distribution $d^2N/d\phi d\eta$ obtained in a NeXSPheRIO
event (presumably also in a true event),
it is not symmetric with respect to the reaction plane.
(We recall that the reaction plane is the plane defined by the impact parameter vector and the beam axis.)
This happens because i) the 
incident nuclei have a granular structure,
ii) the number of produced particles is finite. 
The symmetry might be better with respect to the plane 
with inclination $\psi_2$ in relation to the reference axis and containing the 
beam axis.
Therefore we 
must write for each event
\begin{widetext}
\begin{eqnarray} 
\hspace*{-0.5cm}
\frac{d^2N}{d\phi d\eta}& =& v_0(\eta) [1+ \sum_n 2 v^b_n(\eta) \cos(n(\phi-\phi_b))+ 
\sum_n 2 v'^{b}_n(\eta) \sin(n(\phi-\phi_b)) ]\\
& = & v_0(\eta) [1+ \sum_n 2 v^{obs}_n(\eta) \cos(n(\phi-\psi_2))+ 
\sum_n 2 v'^{obs}_n(\eta) \sin(n(\phi-\psi_2)) ]
\end{eqnarray}
\end{widetext}
It follows that
\begin{equation}
v_2^{obs}(\eta)=v_2^b(\eta) \cos[2(\psi_2-\phi_b)] 
+ v'^{b}_2(\eta) \sin[2(\psi_2-\phi_b)]
\end{equation}
We see that due to the sine term, 
we can indeed have
$<v_2^{obs}(\eta)>><v_2^b(\eta)>$, and therefore $<v_2^{rec}(\eta)>><v_2^b(\eta)>$
 as in  figure 1. The sine term does not vanish upon averaging on events because if a choice such as eq. (4) is done for $\psi_2$,
$v'^{b}_2(\eta)$ and $\sin(2(\psi_2-\phi_b))$ have the same sign. 
This can be visualized with fig. 2a. If the momentum distribution,
instead of being symmetric
with respect to the reaction plane, (for example $v_2^b> 0, v'^{b}_2=0$)
has a positive sine term added ($v'^b_2>0$), it now points at an angle
between 0 and $\pi/4$ above the reaction plane. This angle is in fact $\psi_2$ and is determined experimentally with eq. (4). Therefore $v'^b_2\,\sin(2(\psi_2-\phi_b))>0$. Similarly, if $v'^b_2<0$, $\psi_2$ is between $-\pi/4$ and 0 and
$v'^b_2\,\sin(2(\psi_2-\phi_b))>0$.
(Rigorously,
this sign condition
is true if $\psi_2$ is computed for the same $\eta$ as $v'^b_2(\eta)$.
Due to the actual way of experimentally extracting $\psi_2$, 
we expect this condition is approximately satisfied only 
for particles with small  or moderate pseudorapidity, which are close enough to where $\psi_2$ was computed.)
\begin{center}
\begin{figure}
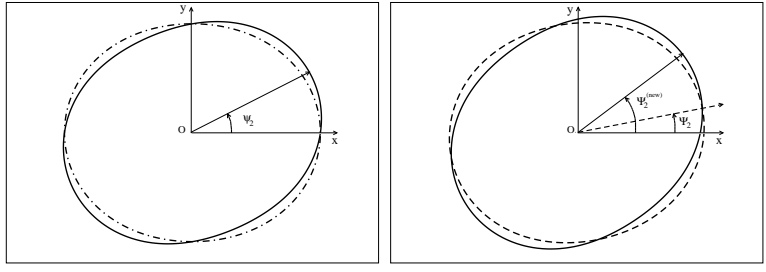

\epsfig{file=fig2top.eps,height=3.5cm,angle=0}
\epsfig{file=fig2bot.eps,height=3.5cm,angle=0}
\caption[]{Assuming (top)
$d^2N/d\phi\,d\eta=1+2v_2^b\cos(2(\phi-\phi_b))
+2v'^b_2\sin(2(\phi-\phi_b))$ with $\phi_b=0$: dash-dotted 
 momentum distribution is symmetric with respect to the reaction plane ($v_2^b>0,v'^b_2=0$) and solid is asymmetric ($v_2^b>0,v'^b_2>0$);
assuming (bottom) $d^2N/d\phi\,d\eta=1+2v_2^{obs}\cos(2(\phi-\psi_2))
+2v'^{obs}_2\sin(2(\phi-\psi_2))$ with $\phi_b=0$: dashed
momentum distribution is symmetric with  to the plane inclination
 $\psi_2$ above the impact parameter and containing the beam axis
($v_2^{obs}>0,v'^{obs}_2=0$) and solid is asymmetric ($v_2^{obs}>0,
v'^{obs}_2>0$).}  
\end{figure}
\end{center}

In the standard approach, for example as  in Phobos analysis, it is {\em assumed} that 
$d^2N/d\phi d\eta$  is symmetric with respect to the reaction plane
and 
there are no sine terms in the Fourier
decomposition in (eq. (5)); 
eq. (7)  leads to (for the hit-based or track-based method)
\begin{equation}
 <v_2^b(\eta)>=<v_2^{obs}(\eta)>/<\cos[2(\psi_2-\phi_b)]>
\end{equation}
Then using $<cos[2(\psi_2-\phi_b)]>=<cos[2(\psi_2^{>}-\phi_b)]>
=<cos[2(\psi_2^{<}-\phi_b)]>$
and 
$<cos[2(\psi_2^{>}-\psi_2^{<})]>=<cos[2(\psi_2^{>}-\phi_b)]><cos[2(\psi_2^{<}-\phi_b)]>=<cos[2(\psi_2^{>}-\phi_b)]>^2$ (where it
is assumed
that the distributions of $\psi_2^{>}-\phi_b$ and $\psi_2^{<}-\phi_b$
are symmetrical with respect to the reference axis and 
$\psi_2^{>}-\phi_b$ and $\psi_2^{<}-\phi_b$ are independent),
eq. (2) follows.
However as explained above, the
use of
NeXus 
initial conditions  leads to $d^2N/d\phi d\eta$ 
not symmetric 
with respect to the reaction plane (and presumably this is also the case in each real event), so eq. (8) and (2) are
 not valid.

As already mentioned, 
the symmetry might be better with respect  to the plane 
with inclination $\psi_2$ in relation to the reference axis and containing the 
beam axis.
From (5) and (6), we have
\begin{equation}
 v_2^b(\eta)=v_2^{obs}(\eta)\times \cos[2(\psi_2-\phi_b)]
+ v'^{obs}_2(\eta)\times \sin[2(\psi_2-\phi_b)].
\end{equation}
If the symmetry is perfect $v'^{obs}_2=0$. Otherwise,
looking at fig. 2b, 
if the angular distribution,
instead of being symmetric
with respect to the axis with inclination $\psi_2$ in relation to the impact parameter, (for example $v_2^{obs}> 0, v'^{obs}_2=0$)
has a positive sine term added ($v'^{obs}_2>0$), it now points at an angle
$\psi_2^{new}$
greater than $\psi_2$. If a negative sine term is added ($v'^{obs}_2<0$), it now points at an angle $\psi_2^{new}$
smaller than $\psi_2$. Both possibilities are equally likely
for a given $\psi_2$ but lead to opposite signs for 
$v'^{obs}_2(\eta)\times \sin[2(\psi_2^{new}-\phi_b)$ (in general). 
Therefore $<v'^{obs}_2(\eta)\times \sin[2(\psi_2-\phi_b)]>=0$. So 
whether the symmetry is perfect or approximate,
$<v_2^b(\eta)>\sim <v_2^{obs}(\eta)\times \cos[2(\psi_2-\phi_b)]>$ and
instead of eq. (2) we would have
\begin{equation}
<v_2^{Rec}(\eta)>=<v_2^{obs}(\eta)\times
         \sqrt{<\cos[2(\psi_2^{<0}-\psi_2^{>0})]>}>
\end{equation}
In figure 3, we show $<v_2^{Rec}(\eta)>$ (dash-dotted line)
and $<v_2^b(\eta)>$ (solid line). We see that 
the agreement between both methods is improved compared to figure 1. 
\begin{center}
\begin{figure}
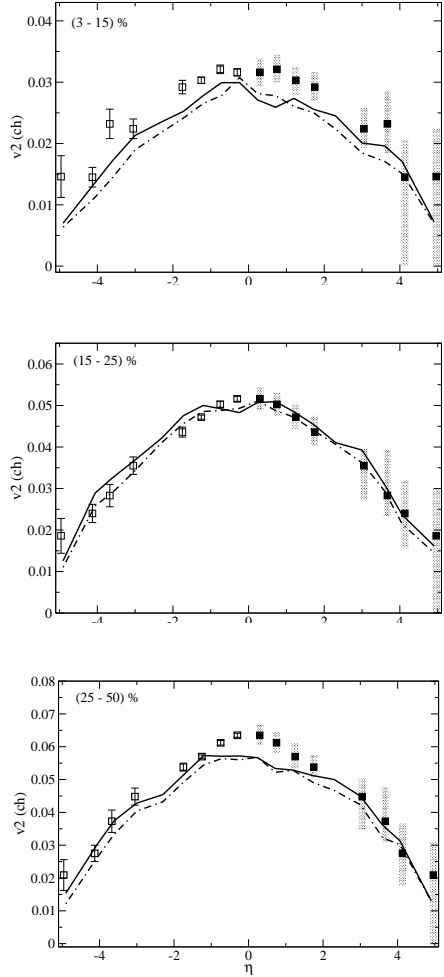

\epsfig{file=fig3top.eps,height=4.0cm}\\
\vspace*{0.5cm}
\epsfig{file=fig3med.eps,height=4.0cm}\\
\vspace*{0.4cm}
\epsfig{file=fig3bot.eps,height=4.0cm}
\caption[]{Comparison between true elliptic flow $v_2^b$  (solid line) and 
suggested method to compute reconstructed elliptic flow from data $v_2^{Rec}$
(dash-dotted)
 for the three Phobos centrality windows\cite{phobos}.
Squares represent Phobos data (black error bars are 1 $\sigma$ statistical
errors and grey bands, systematic uncertainties at $\sim$90\% confidence level).
}  
\end{figure}
\end{center}
We have also computed  the elliptic flow parameter as function of transverse momentum 
for charged hadrons with $0<\eta<1.5$ for the 50\% most central collisions.
We found that 
$<v_2^b(p_\perp)>$  computed as in eq. (1) is well approximated by
$<v_2^{Rec}(p_\perp)>$  computed as in eq. (10).

In summary, from figure 1, elliptic flow estimated from the standard method with reaction plane correction is an overestimate of true elliptic flow
($v_2^{rec}>v_2^b$). 
From figure 3,
using a method that takes into account the more symmetrical nature of particle distribution in relation to the plane with inclination $\psi_2$ with respect to the reference axis and containing the beam axis
(rather than with respect to the reaction plane), we get
a better agreement between reconstructed and true elliptic flows 
($v_2^{Rec}\sim v_2^b$).

As for overestimating the true elliptic flow, a similar conclusion was reached in
\cite{miller} and \cite{zhu}. In \cite{miller},
elliptic flow was assumed proportional to eccentricity and eccentricity was computed event-by-event using a Monte Carlo Glauber calculation. As in our case, $\vec{b}$ is known. It was found that the integrated true $v_2^b$ is smaller than $v_2^{rec}$ computed with a two-particle cumulant method 
(for all centralities) and larger than $v_2^{rec}$ computed with higher order cumulants
(for centralities 0-80\%).
In \cite{zhu}, elliptic flow was computed event-by-event within the UrQMD model. Again $\vec{b}$ is known. It was found that 
the integrated true $v_2^b$ is smaller than $v_2^{rec}$ computed with a two-particle cumulant method 
(for all centralities) and equal to $v_2^{rec}$ computed with higher order cumulants
(for centralities 10-50\%). Differential elliptic flow was also studied
leading to similar conclusions.

In these two papers, it is expected \cite{miller,zhu}
that there will be differences between  $v_2^b$
and $v_2^{rec}$ calculated with the reaction plane method or two-particle cumulant method both because of the so-called
non-flow correlations (overall momentum conservation, resonance decays, jet production,etc) and event-by-event fluctuations (mostly eccentricity fluctuations).
 In principle, higher-order cumulant methods  take care of non-flow 
effects.
If there is still disagreement between the true elliptic flow and 
higher-order cumulant methods, as in  \cite{miller},
then fluctuations are important. 
If there is agreement as in  \cite{zhu}, 
then non-flow effects are important and not fluctuations. In addition to
 the disagreement between their conclusions, 
\cite{miller} and \cite{zhu}
do not (neither are expected to)
reproduce the RHIC data. 
So an interesting question is whether a  more accepted 
hydrodynamical description would lead to a sizable effect. Using NeXSPheRIO,
we found that true elliptic flow $v_2^b(\eta=0)$ is overestimated by 
$\sim$ 15-30 \% (according to 
centrality) with the reaction plane method, and $v_2^b(p_\perp)$ by $\sim$ 30\% at $p_\perp$=0.5 GeV.
 In our case, since $<v_2^b> \sim <v_2^{Rec}>$,
a large part of the difference between the true
$<v_2^b>$ and reconstructed $<v_2^{rec}>$ is due 
to the (wrong) assumption of symmetry of the particle distribution around the reaction plane, made to obtain $<v_2^{rec}>$.

Finally, we would like to emphasize that it is important to have precise
experimental determination of elliptic flow, in particular free from the assumption of symmetry that we discussed.
Elliptic flow teaches us about the initial conditions and thermalization, in 
principle. In this manner, in \cite{Hirano1}, the author showed that with 
his hydrodynamical code plus freeze out, he could not reproduce $v_2(\eta)$,
in particular at large $\eta$, and therefore concluded that there might be a lack of thermalization for these large $\eta$'s. In \cite{Hirano2}, 
it was shown that agreement with $v_2(\eta)$
data could be obtained for central collisions
with a similar hydrodynamical code but with
color glass initial conditions and,
instead of freeze out, a transport code 
matched to the hydrodynamical code
to describe particle emission. It was therefore concluded that some 
viscosity was necessary in the hadronic phase.
Lastly, in \cite{Hirano3} (see figures 3 and 4), it was shown that 
for all centralities, Glauber-type initial conditions plus hadronic dissipation
lead to a reasonable agreement with $v_2(\eta)$ data
while color glass condensate initial conditions plus hadronic dissipation do 
not,  except in the most central window
(unless some additional dissipation occurs in the early quark gluon plasma phase). 
Both sets of initial conditions without hadronic dissipation tend to underestimate 
$v_2(\eta)$ data if $T_{f.out}=169$MeV and overpredict them if $T_{f.out}=100$MeV.
However  these conclusions would be affected if $v_2(\eta)$ data were lower, as we think they should be. (Incidently,
though our objective was not to reproduce data, note that our model with freeze out (no transport code)
reproduces reasonably both the $v_2(\eta)$ data as in 
\cite{Hirano3} (figure 3) and the $v_2(p_\perp)$ data (not shown).)
Therefore, to know e.g. what the initial conditions are or if there is 
viscosity and in what phase, we need to settle the question of whether 
event-by-event fluctuations 
are important and take them into account in the experimental analysis.

We acknowledge financial support by FAPESP (2004/10619-9, 2004/13309-0,
2004/15560-2, CAPES/PROBRAL, CNPq, FAPERJ and PRONEX.

\end{document}